# Thermal boundary conductance of CVD-grown MoS$_2$ monolayer-on-silica substrate determined by scanning thermal microscopy

C.M. Frausto-Avila[a,b], V. Arellano-Arreola[b], J.M. Yañez-Limon[b], A. De Luna-Bugallo[b,c,*], S. Gomès[a], P.-O. Chapuis[a,*]

[a] Univ Lyon, CNRS, INSA-Lyon, Université Claude Bernard Lyon 1, CETHIL UMR5008, F-69621, Villeurbanne, France
[b] Cinvestav Unidad Querétaro, Querétaro, Qro. 76230, Mexico
[c] Departamento de Nanotecnología, Centro de Física Aplicada y Tecnología Avanzada, Universidad Nacional Autónoma de México, CP 76000, Querétaro, Qro., Mexico.
*Corresponding authors: aluna@fata.unam.mx, olivier.chapuis@insa-lyon.fr

We characterize heat dissipation of supported MoS$_2$ monolayers grown by chemical vapour deposition (CVD) by means of ambient-condition scanning thermal microscopy (SThM). We find that the thermal boundary conductance of the MoS$_2$ monolayers in contact with 300 nm of SiO$_2$ is around $4.6 \pm 2$ MW.m$^{-2}$.K$^{-1}$. This value is in the low range of the values determined for exfoliated flakes with other techniques such as Raman thermometry, which span an order of magnitude (0.44 - 50 MW.m$^{-2}$.K$^{-1}$) and underlines the dispersion of the measurements. The sensitivity to the in-plane thermal conductivity of supported MoS$_2$ is very low, highlighting that the thermal boundary conductance is the key driver of heat dissipation for the MoS$_2$ monolayer when it is not suspended. In addition, this work also demonstrates that SThM calibration using different thicknesses of SiO$_2$, generally employed in bulk materials, can be extended to 2D materials.

Over the last two decades, there has been a growing interest in 2D materials due to their low dimensionality, making them attractive for various fields such as electronics, condensed matter, photonics, catalysis, among others. After the popularization of graphene, different layered materials have been discovered, including borophene[1], hexagonal boron nitride hBN[2], and transition metal dichalcogenides[3] (TMDCs). Molybdenum disulfide (MoS$_2$), a member of the family of TMDCs, is a semiconductor whose synthesis has been quite well developed and established by different approaches[4–9]. In the case of a single layer (thickness around 6 Å) MoS$_2$ exhibits a direct bandgap[10] (≈1.82 eV), reasonable electrical conductivity, large spin-orbit coupling, strong exciton binding, which makes it suitable for several optoelectronic applications[11,12].

Investigating the properties of 2D materials and implementing various characterization techniques are challenging in many cases, particularly for thermal studies, due to the complexity associated with their extremely-low thickness. With the advent of atomically thin materials, different thermal characterization techniques have been extrapolated from bulk to nanostructured materials. Techniques such as the 3ω method[13], photothermal characterization[14], as well as Raman thermometry[15], have been efficiently translated for thermal conductivity measurements of such materials. Some of the techniques require depositing metallic contacts onto the samples[16], which is unfeasible for certain configurations of the systems, or high-frequency equipment[17] with assumptions on the (ideal) optical absorption.

Thermal characterization aims mainly at obtaining parameters such as thermal conductivity and thermal boundary conductances (TBCs). Due to the quick preparation and crystal quality, most of the reports regarding the thermal properties of MoS$_2$ are normally performed using exfoliated samples (either supported by an arbitrary substrate or suspended in a micrometer-sized hole), however MoS$_2$ crystals can also be grown by chemical vapor deposition (CVD), which appears more appropriate for device integration and scaling[18]. For supported exfoliated monolayers (flakes), thermal conductivity values in the range 34.5-62.0 W.m$^{-1}$.K$^{-1}$ are reported, while TBCs span 0.44-50 MW.m$^{-2}$.K$^{-1}$ [19–22]. For the suspended configuration, thermal conductivity values between 23.2 and 84.0 W.m$^{-1}$.K$^{-1}$ are reported[21,23,24]. These values are more accurate when averaged over large areas and were obtained by techniques with inherent limitations such as optical diffraction[25] in the best cases. Beyond such scales scanning thermal microscopy (SThM), developed since the 1990s[25] on the atomic force microscopy (AFM) platform, is attractive since the spatial resolution can depend only on the radius of the thermal contact between the probe and the sample. Such radius can reach the sub-100 nm scale under certain operation conditions, making it an option for nanoscale thermal measurements, in particular thermometry[26]. Although SThM was already used on structures involving MoS$_2$ for thermometry in complex devices[26] and for an analysis of heat dissipation in samples where MoS$_2$ was coupled to graphene[27], it has not been used for quantitative thermal-property determination of the TMDC yet. In the present work, we propose a methodology based on ambient-condition SThM to determine the TBC value for MoS$_2$ monolayers grown by CVD on SiO$_2$/Si substrates. It is demonstrated that (in-plane) thermal conductivity is not useful in practice for samples with several microns of lateral lengths, since heat dissipation takes place towards the substrate.

The MoS$_2$ crystals are grown by atmospheric CVD, further details can be found in previous reports[28]. The studied systems are composed of a MoS$_2$ monolayer supported by a 300 nm-thick silica layer standing over a silicon wafer. Figure 1 shows an optical image of a typical MoS$_2$ monolayer, and the overall stack is reminded in the insert. The typical lateral size of the crystals is around 70-100 µm. As a large number of MoS$_2$ monolayers crystals (typical shapes as that of Fig. 1) can be present on the substrate, careful attention is paid to avoid thermal or optical crosstalk. Moreover, we use Raman spectroscopy to monitor the frequency difference between the $E_{2g}$ and $A_{1g}$ peaks[7] to select only the single-layered MoS$_2$ crystals (see Suppl. Fig. 1).

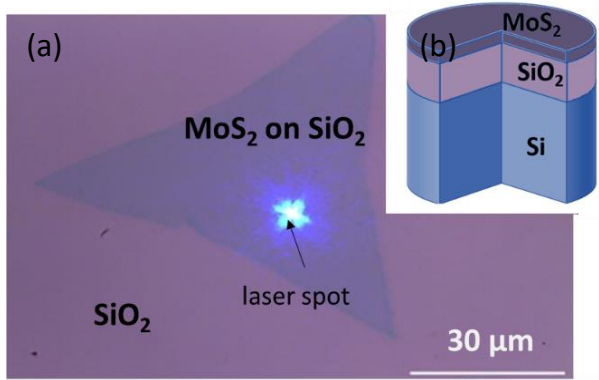

Figure 1. (a) Optical microscopy image of a MoS$_2$ monolayer with a triangular shape. The light dot corresponds to a laser spot irradiating the surface. (b) Cross-section schematic of the analyzed system.

Thermal scans are acquired by means of thermoresistive SThM, with two different thermal probes[25]. The data reported here are obtained using a Wollaston probe, whose sensor is a 5 µm-in-diameter Pt$_{90}$/Rh$_{10}$ filament with a length of ≈200 µm. It is bent in a V shape with the tip contacting the sample, and anchored between unetched parts of the Wollaston wire, where the Pt$_{90}$/Rh$_{10}$ alloy is surrounded by a silver shell (≈75 µm of diameter in total). This makes the sides of the filament less electrically-resistive. As a consequence, the filament is self-heated when fed by an electrical current $I$. The SThM operation mode typically used in this work consists in bringing the heated probe into contact with the sample in order to heat it locally and scanning its surface at constant force as in AFM. Noticeably, the electrical resistance of the sensor $R$ depends on the average temperature of the probe $\bar{T}$, so in addition to being a heat source the probe is a thermometer:

$$R = R_0 \cdot (1 + \alpha\theta), \quad (1)$$

where $\alpha = (1/R) \cdot dR/d\bar{T}$ is the temperature coefficient of the Pt$_{90}$/Rh$_{10}$ electrical resistance known to be $1.66 \times 10^{-3}$ K$^{-1}$, $R_0 = R(T_0)$ and $\theta = \bar{T} - T_0$ is the temperature rise above ambient temperature $T_0$ (see Suppl. Sec. 2 for more details on SThM). The ratio between the heat input to the sensor $P = RI^2$ and the sensor average temperature rise $\theta$ provides a qualitative estimation of the sample ability to dissipate heat, it is known as the probe thermal conductance $G_{\text{probe}}$ (see Suppl. Secs. 2-4) and its value is close to 95 µW.K$^{-1}$ in ambient condition.

In our setup, the current supplied to the probe is constant, and the voltage variation $\Delta V$ is monitored at the same time than the topography during the scan[25] of the sample surface. The voltage reference ($\Delta V = 0$) is taken at an arbitrary point on the surface. Note that thermal stabilization is reached by waiting around 45 min before scanning to minimize the impact of thermal drifts on the images. Figure 2 displays (a) the recorded AFM topography, and (b) the raw thermal image ($\Delta V$) of a MoS$_2$ monolayer. It is possible to directly correlate the crystal topography (here slightly different from the crystal of Fig. 1) with the thermal contrast. One can notice the strong difference between the thermal signal on the MoS$_2$ monolayer and that in the region around (SiO$_2$/Si substrate). Artifacts linked to scan direction are observed in the topography image, and are also present in the raw thermal image.

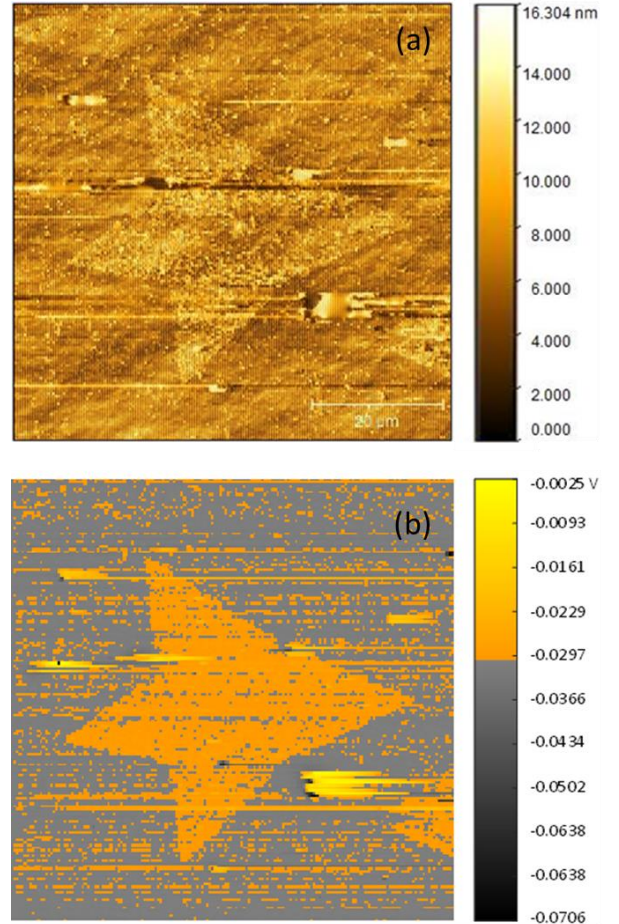

Fig 2. (a) Topography image obtained by atomic force microscopy with a Wollaston probe. A flat plane was subtracted from the raw image. (b) Raw thermal signal ($\Delta V$) obtained during the same scan.

The raw thermal image can be translated into a probe average temperature image with Eq. (1). One obtains the probe temperature variation $\Delta\theta$ as a function of location on the sample (with respect to some reference, here arbitrarily taken as the lowest value of the image). In order to smoothen the thermal signal fluctuations, we average the signal close to an edge as shown in Fig. 3 (the image is rotated with respect to that of Fig. 2). It is found that the probe temperature increases by approximately 0.1 K when it moves from SiO$_2$ to MoS$_2$, indicating that the MoS$_2$ layer induces an additional thermal resistance for the flux being dissipated into the sample. At first sight, this effect could be ascribed either to a worse contact between the SThM probe with MoS$_2$ than silica or to a weak thermal contact between MoS$_2$ and the silica. This is in striking contrast to supported graphene, which increases heat dissipation properties[29,30].

The temperature map is then translated into a map of the probe thermal conductance $G_{\text{probe}}$ (again with respect to an arbitrary reference, see Suppl. Sec. 2). It is found that $G_{\text{probe}}$ varies by $\Delta G_{\text{probe}} = 55 \times 10^{-9}$ W.K$^{-1}$ close to the edge of the MoS$_2$ crystal. It is instructive to compare this value with that obtained when simply increasing the thickness of the silica layer (silica is a standard solid-state thermal insulator). In Ref. [31], some of us reported, with a similar Wollaston SThM probe, how $G_{\text{probe}}$ varies with SiO$_2$ thickness (see Suppl. Sec. 5). Assuming similar thermal conductivity for the oxide in the SiO$_2$/Si substrate here and that of Ref. [31], we find that the decrease of probe thermal conductance when locating the probe on MoS$_2$ is the same as that while bringing it over an oxide layer thicker by 95 nm. This thickness is more than hundred times than that of MoS$_2$, underlining the potential of the TMDC as thin but efficient heat barrier.

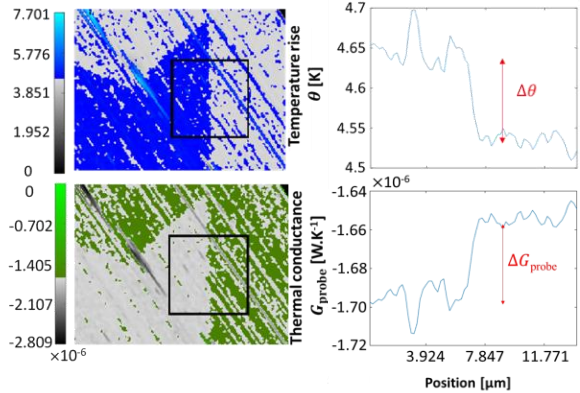

Figure 3. (Top) Probe temperature rise with respect to an arbitrary reference between the supported MoS$_2$ and the silica-over-silicon wafer. (Down) Probe thermal conductance deduced from the temperature measurements (arbitrary reference). Left panels show rotated images with respect to Fig. 2, and right ones vertical averages in the square for each (horizontal) position.

In the following, we aim at obtaining quantitative thermal data for the MoS$_2$ monolayer (see Suppl. Sec. 3 for a graphical summary of the procedure). To determine these, one needs first to find an estimate of the thermal contact radius $b$, i.e. the size over which the SThM probe heats the sample. It is obtained by first comparing the probe thermal conductance with that obtained as a function of the silica thickness in Ref. [31]. The effective thermal conductivity (that of a bulk leading to the same $G_{\text{probe}}$) determined for a layer of 300 nm of SiO$_2$ over Si is around $\lambda_{eff} \approx 2$ W.m$^{-2}$.K$^{-1}$ (see Suppl. Sec. 6). The radius can then be obtained from a finite element (FE) simulation solving the steady-state heat equation, in the sample only. Indeed, the sample thermal conductance (conductance associated with heat dissipation in the sample from a hot isothermal disc on the sample surface) is $4\lambda_{eff}b$ and equal to that of the exact geometry (300 nm SiO$_2$/Si) for an identical thermal contact radius. The radius determined from the FE simulation is around 4 µm (see Suppl. Sec. 7 for more details). This value underlines the well-known fact that heat spreads from the probe to the sample in the air, leading to a transfer over a much larger area than that of the mechanical contact[25]. Since heat is transferred mostly through air to the sample, the thermal boundary conductance at the mechanical contact is not a matter of concern. Note that the impact of the thermal contact conductance between the tip and the sample depends only on the effective (bulk) thermal conductivity felt by the probe for heat transfer through air[32]. This is in stark contrast to many works where heat transfer inside the whole system made of the SThM sensor and at the probe-sample contact is also required to be modelled. This simplification is possible because the current work builds on the previous calibration in Ref. [31].

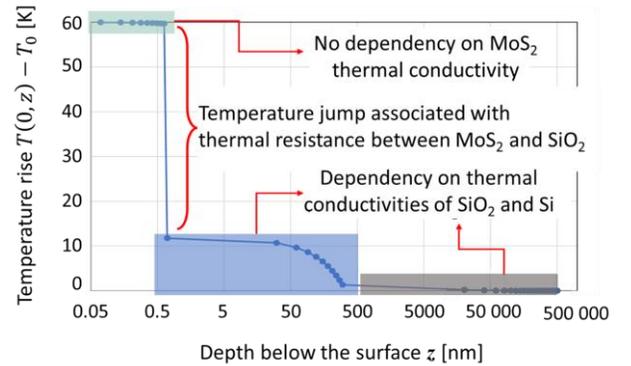

Figure 4. Temperature profile in logarithmic scale as a function of depth below the heat source (FEM simulation). The upper region overlaid in green corresponds to the MoS$_2$ monolayer, the lower blue region corresponds to SiO$_2$, and, finally, the gray region corresponds to the contribution of the Si substrate.

The final step is also performed with a FE simulation. The actual geometry, i.e. the stack shown in the inset of Fig. 1, is considered, with known thermal conductivity values for silicon and silica, and again with a disc of homogeneous temperature as heat input on the top (see Suppl. Sec. 8). The unknowns are the MoS$_2$ thermal conductivity, supposed isotropic in the 0.7 nm thickness, and the TBC between MoS$_2$ and silica. These two quantities are adjusted in the 2D cylindrical FE simulation to dissipate a power equivalent to that of a bulk with the effective thermal conductivity mentioned above (i.e. the bulk and $MoS_2/SiO_2/Si$ sample thermal conductances are equal). It is found that the value of the thermal conductivity of $MoS_2$ impacts very weakly the temperature distribution, which is driven only by the TBC. The temperature profile in the center of the structure is provided in Fig. 4 as a function of depth. Note that we verified that the $MoS_2$ lateral size and shape do not matter provided that the size is larger than the thermal contact radius. The temperature profile is mostly flat in the thin TMDC layer (see Suppl. Fig. 8 for 3D temperature distribution), as a result of the insensitivity to thermal conductivity. Most importantly, there is a strong temperature discontinuity associated with the $MoS_2/SiO_2$ interface. Finally, one obtains a value of 4.6 ± 2 $MW.m^{-2}.K^{-1}$ for the thermal boundary conductance, which is close to the values found experimentally for flakes by Raman thermometry[12,20] and of similar order of magnitude to a molecular dynamic study[33]. The value is intrinsically low as Van Der Waals bonding provides a much weaker connection between the monolayer and its support. It seems therefore that the quality of the material, being it an exfoliated flake or a CVD-grown crystal, is not key for heat conduction when it is supported.

In summary, this works has showed that, with a proper calibration technique, SThM allows for quantitative determination of key parameters associated with heat dissipation in supported 2D materials. Thermal conductivity may not be the relevant parameter, while Van Der Waals bonding leads to weak thermal coupling with substrates. In the near future, it will be useful to analyze TMDCs with a better spatial resolution, either by studying the jump at contact in probe approach curves or by implementing vacuum conditions. Analyzing heat dissipation in TMDCs as a function of temperature may also enable to discriminate between the effect of thermal conductivity and thermal boundary conductance[27].

See the Supplementary Information for details on material, the temperature-probe thermal conductance connection, for details on the varying-thickness oxide calibration samples and on the simulations.

We thank J.-M. Bluet for helping with the Raman characterization and A. Alkurdi for discussions. C.M.F.A. acknowledges a CONACYT scholarship funding. S.G. and P.-O.C. acknowledge support of project TIPTOP (ANR-16-CE09-0023) and A.D.L.B. the support of project UNAM-PAPIIT-A101822.

# SUPPLEMENTARY MATERIAL TO

# Thermal boundary conductance of CVD-grown MoS₂ monolayer-on-silica substrate determined by scanning thermal microscopy

C.M. Frausto-Avila[a,b], V. Arellano-Arreola[b], J.M. Yañez-Limon[b], A. De Luna-Bugallo[b], S. Gomès[a], P.-O. Chapuis[a]

[a] Univ Lyon, CNRS, INSA-Lyon, Université Claude Bernard Lyon 1, CETHIL UMR5008, F-69621, Villeurbanne, France.
[b] Cinvestav Unidad Querétaro, Querétaro, Qro. 76230, Mexico

**Contents**

1. *Procedure to determine if the sample is a MoS₂ monolayer*
2. *Procedure to determine the probe thermal conductance*
3. *Brief summary of the different steps for the data treatment*
4. *Thermal circuit associated to heat dissipation from the probe*
5. *'Equivalent oxide thickness' procedure*
6. *Equivalent effective thermal conductivity*
7. *Determination of the thermal contact radius*
8. *Determination of the thermal boundary conductance*

## 1. Procedure to determine if the sample is a MoS₂ monolayer

According to the methodology followed by Ganorkar *et al.*[1] for estimating the number of MoS₂ layers synthesized by Chemical Vapor Deposition (CVD), the difference between the Raman shifts of the $E_{2g}$ and $A_{1g}$ peaks is used to estimate the number of layers. The wavenumber difference is $\Delta\sigma = 21.5$ cm$^{-1}$ for a monolayer and $\Delta\sigma = 22.3$ cm$^{-1}$ for bilayers [1]. In this work we find a value of $\Delta\sigma = 21.6$ cm$^{-1}$, which can safely be considered as a MoS₂ monolayer crystal (see Suppl. Fig. 1).

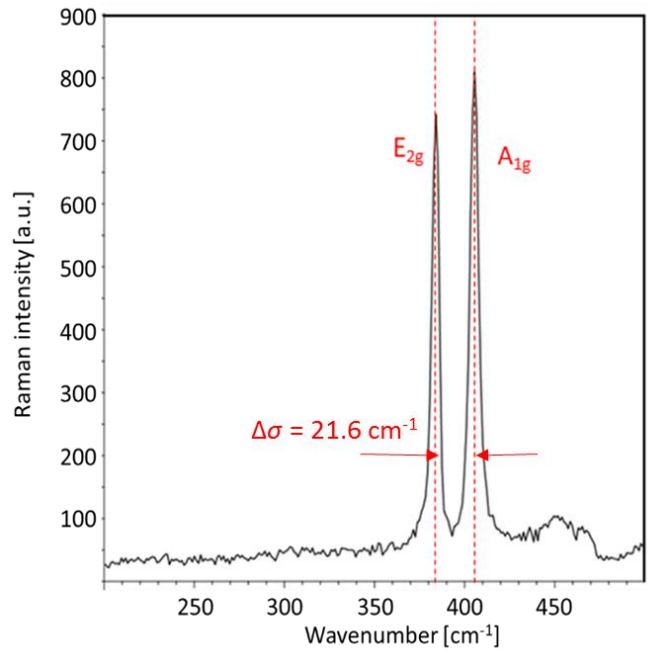

*Suppl. Fig. 1. MoS₂ Raman spectrum.*

## 2. Procedure to determine the probe thermal conductance

The probe thermal conductance is defined as
$$G_{probe} = P/\theta, \tag{1}$$
where $P = R\ I^2$ is the Joule self-heating power inside the sensitive part (sensor) of the probe, $R$ the electrical resistance of the sensitive part of the probe (sensor), $I$ is the electrical current in the probe and $\theta$ its mean temperature rise with respect to ambient. The Wollaston probe resistance is inserted into a Wheatstone bridge (see Suppl. Fig. 2), and it is the bridge imbalance voltage $\Delta V$ that is provided in the thermal image. As a result, scans provide only temperature rises
$$\Delta\theta = \frac{\Delta V}{I} \cdot \frac{1}{\alpha R} \tag{2}$$
relative to an absolute reference temperature $\theta_{ref} + T_0$, which is selected by balancing the bridge ($\Delta V = 0$), and not directly the probe voltage $V_p$. It is customary to balance the bridge either far from contact ($\theta_{ref} = \bar{T} - T_0$, where $\bar{T}$ is the average temperature in the sensor), or in contact at a given location on the sample ($\theta_{ref} = \bar{T} - T_{ref}$). Here, the second option is chosen.

Knowing the value of the electrical resistance in the bridge $R_v$, one can deduce the probe temperature $T_0 + \theta = T_0 + \theta_{ref} + \Delta\theta$. We use a symmetric bridge, so that $R_1 = R_2$ (taken as 50 $\Omega$), and an input bridge current of $I_{in} = 80$ mA, i.e. $I = 40$ mA is supplied in the probe. The electrical resistance of the sensitive part of the Wollaston probe ($Pt_{90}/Rh_{10}$ filament) $R$ is computed by determining the geometrical parameters, noticing that the variable resistance, when the bridge is balanced, is equal to:
$$R_v = R_{wiring} + R_{Ag} + R\ , \tag{3}$$
where the electrical resistance of the wiring is estimated to be $R_{wire} \approx 1\ \Omega$, $R = R_0(1 + \alpha\ (\theta))$ and $R_{Ag}$ is obtained by subtraction from room-temperature measurements ($\theta = 0$). $R_{Ag}$ is the electrical resistance of the Wollaston wire (cantilever) assumed made of the silver shell.

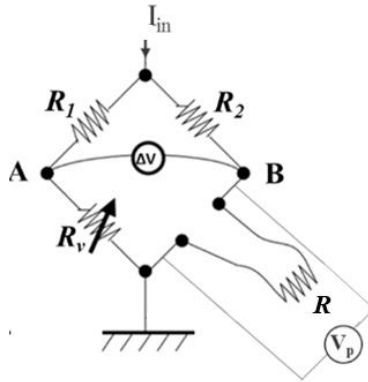

*Suppl. Fig. 2. Wheatstone bridge where $R_1$ and $R_2$, $R_v$ and $R$ are the fixed electrical resistances, the variable electrical resistance and the electrical resistance of the probe respectively.*

The local probe thermal conductance variation is obtained as in Ref. [2] by differentiating logarithmically Eq. (1), which gives after straightforward algebra:
$$\Delta G_{probe} = G_{probe} - G_{probe\ ref} = G_{probe\ ref} \cdot \Delta\theta \cdot \left[\alpha(\theta) - \frac{1}{\theta}\right] \tag{4}$$
if the current variation in the probe $\Delta I$ is neglected (which we verified). $G_{probe\ ref}$ is the probe thermal conductance at absolute temperature $T_0 + \theta_{ref}$. Of course Eq. (4) is valid only provided the thermal conductance variations stay small. A direct calculation without linearization can be performed if this is not the case.

From the parameters experimentally determined, $G_{probe\ 0} \approx 95\ \mu W.K^{-1}$ and $\theta_{ffc} \approx 156$ K far from contact. When the probe contacts the sample, the temperature rise $\theta$ decreases by about 10% for materials of moderate thermal conductivities, so $\theta_{ref} \approx 140$ K. In principle $G_{probe\ ref}$ and $G_{probe\ 0}$ are different, but in the following Eq. (4) is used with the assumption $G_{probe\ ref} \approx G_{probe\ 0}$, which induces an uncertainty propagation in $\Delta G_{probe}$. Note that maps of $\Delta G_{probe}$ or $G_{probe}$ with respect to an arbitrary reference provide similar information.

## 3. Brief summary of the different steps for the data treatment

The raw SThM image allows only acquiring qualitative analysis of heat dissipation at the sample surface and a significant part of the work is therefore to deduce quantitative data from these images. We summarize the different steps (see Suppl. Fig. 3) mentioned in the main manuscript here, and more details are provided in the Sections below.

- First (1), the local probe thermal conductance $G_{probe}$ is obtained from the electrical data (see above).
- Then (2), the calibration from Guen *et al.* [3] allows obtaining an oxide thickness that impacts the probe thermal conductance equivalently as MoS$_2$. This step is interesting for qualitative reasoning but not decisive for the following.
- More importantly (3), the same paper [3] allows determining the two effective bulk thermal conductivities $k_{eff}$ that provide the same probe thermal conductance as that of the MoS$_2$/SiO$_2$/Si and the SiO$_2$/Si samples, respectively. Noticeably, all the previous steps do not require simulations. But they do not allow to determine the thermal contact radius $b$.
- The simulation steps (4) involve Finite Element (FE) modelling. We proceed in two steps: (4a) we first determine the thermal contact radius $b$, and then (4b) we use it to determine the MoS$_2$ thermal properties. The thermal radius is obtained by equating the thermal conductances of the effective bulk geometry (known to be $G_{sample} = 4k_{eff}b$) and of the exact SiO$_2$/Si geometry. Then for such radius a FE simulation of the MoS$_2$/SiO$_2$/Si stack is performed. The thermal conductivity of the monolayer $k$ and the thermal boundary conductance between the monolayer and the supporting material $G_{TBC}$ are varied in order to match the stack effective thermal conductivity determined in (3).

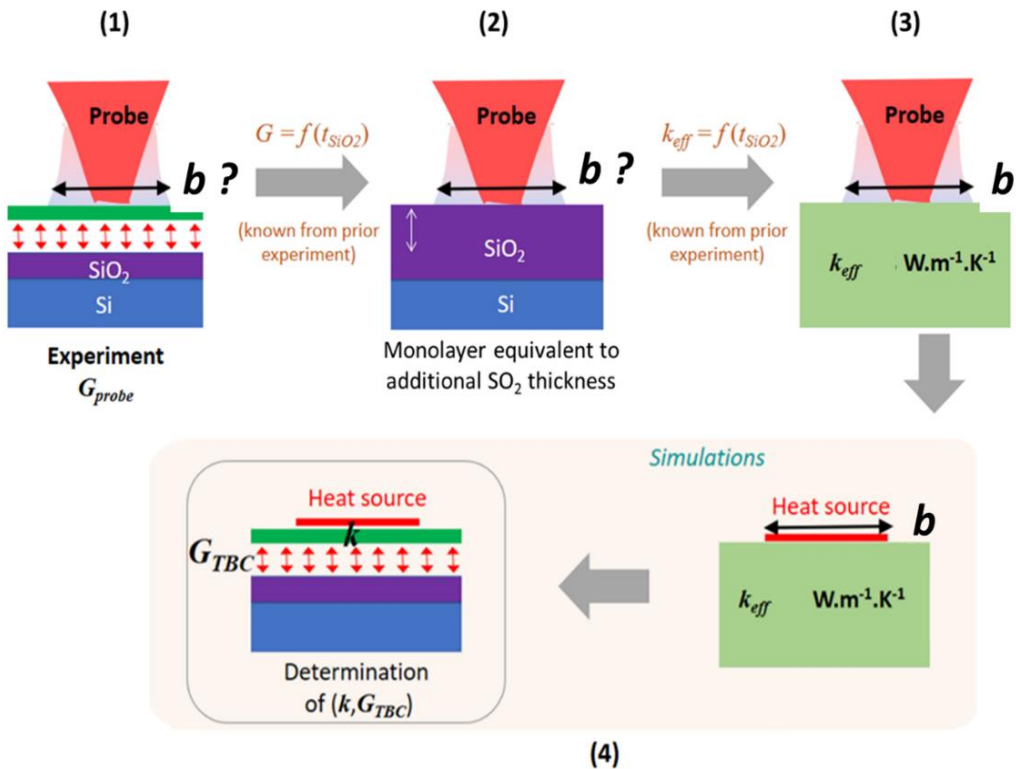

*Suppl. Fig. 3. Schematic of the different steps of our approach for quantitative measurement.*

## 4. Thermal circuit associated to heat dissipation from the probe

We provide a schematic clarifying the different thermal conductances involved in our SThM experiments. $G_{probe}$ is in principle indeed the sum of the three channels allowing heat to dissipate from the probe, where only one is useful for the experiment. However $G_{environement} \approx 0$ when the probe is in contact[3]. $G_{tip}$ includes the thermal contact conductance associated with the transport of heat from the tip into the sample, and is usually difficult to determine precisely. The method described below (Suppl. Sec. 5) avoids addressing this issue fully.

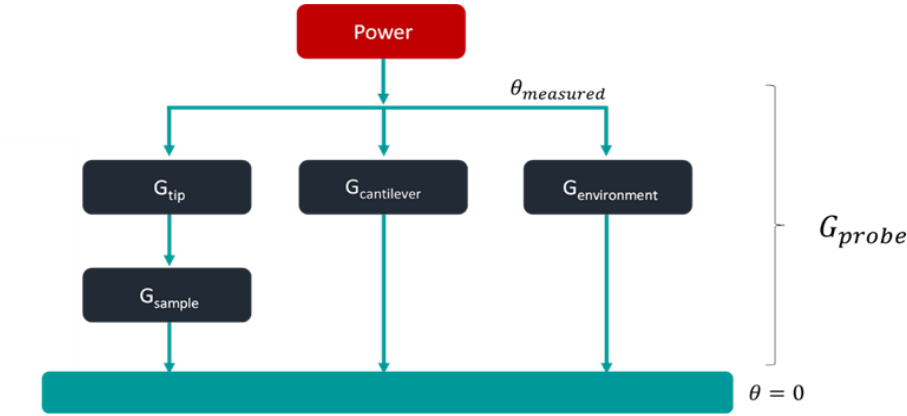

*Suppl. Fig. 4. Suppl. Fig. 3. Thermal circuit associated with heat dissipation in the probe.*

## 5. 'Equivalent oxide thickness' procedure (Step (2) of Suppl. Fig. 3)

To find the equivalent thickness of $SiO_2$ that induces a similar thermal resistance in the sample as that of the monolayer of $MoS_2$, we use a calibration sample[3] made of a mosaic of silicon oxide layers with different thicknesses coating a silicon wafer. It happens that the substrate below $MoS_2$ is similar, with a $SiO_2$ layer on top of the silicon wafer. Since the two samples were not prepared at the same time and with the same goal, some uncertainty is introduced by comparing the data, which propagates until the determination of the thermal boundary conductance. The calibration sample is shown in Suppl. Fig. 5 and detailed in the previous publication[3] (beware that notations are not the same and that the probe thermal conductance defined here is based on the probe average temperature, not the probe apex one).

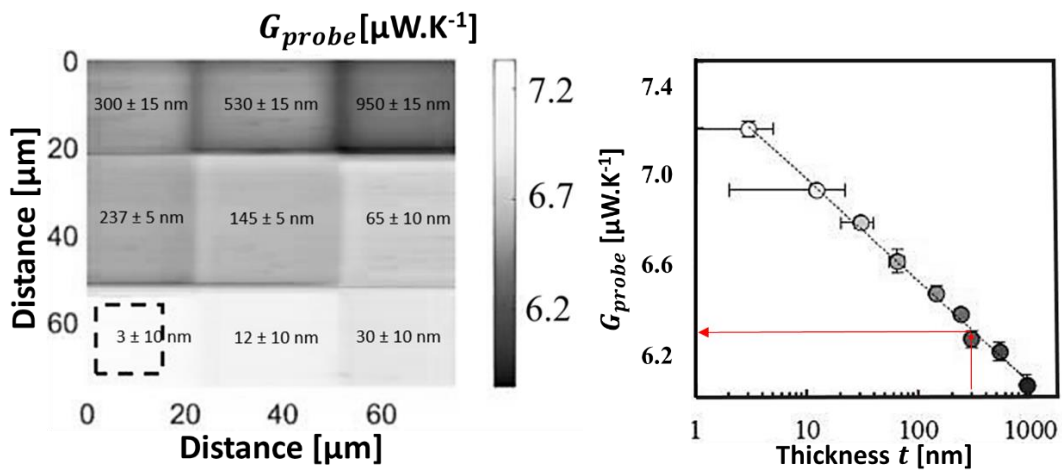

*Suppl. Fig. 5. (Left) SThM scan of the mosaic sample made of 9 different thicknesses of silica. The heights of the silicon oxide steps are indicated on the images. (Right) Probe thermal conductance for the different $SiO_2$ thicknesses. The probe thermal conductance reference ($G_{probe} = 0$) is the far-from-contact position.*

More precisely, $G_\text{probe}$ varies of $\Delta G_{probe} = 55 \times 10^{-9}$ W.K$^{-1}$ when the probe moves from the MoS$_2$ crystal to the oxide surface (see Fig. 3 in the main paper). Note that $\Delta G_{probe} = G_{probe}(\text{MoS}_2) - G_{probe}(\text{SiO}_2\ 300\ \text{nm})$. From Suppl. Fig. 5, we find that $G_{probe}(\text{SiO}_2\ 300\ \text{nm}) + \Delta G_{probe} = G_{probe}(\text{SiO}_2\ 395\text{nm})$ in the calibration sample.

## 6. Equivalent effective thermal conductivity (Step (3) of Suppl. Fig. 3)

The effective thermal conductance for the {oxide on silicon} sample is obtained from the calibration curve in Ref. [3], as shown in Suppl. Fig. 6. The advantage of this method is that it provides a quantity that depends only on the sample and does not require the knowledge of the thermal conductance corresponding to the heat transfer between the probe and the sample included in $G_{tip}$. For the value of $G_{probe}$ found in Suppl. Fig. 5, we find $k_{eff} \approx$ 2.1 W.m$^{-1}$.K$^{-1}$ in Suppl. Fig. 5.

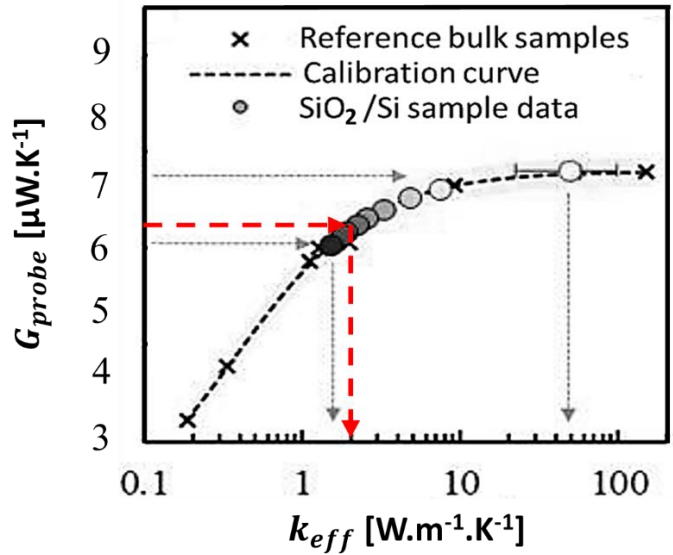

*Suppl. Fig. 6. Variation of the probe thermal conductance (reference far from contact) as a function of the effective thermal conductivity.*

## 7. Determination of the thermal contact radius

The thermal contact radius, i.e. the size of the hot zone on the sample surface (assumed to be a disc), is required for the final step. One can consider that there is a single thermal contact radius for each effective thermal conductivity, i.e. the radius does not depend on the exact configuration within the sample but only on the thermal conductance $G_{sample}$[4]. This conductance is not known initially, and we are required to perform simulations in order determine the thermal contact radius for the {silica on silicon} sample.

With finite-element (FE) simulations, we compute the thermal conductance of a medium consisting of a silica layer (300 nm) over a silicon wafer, for a given radius (see Suppl. Fig. 7). This can be done for an arbitrary temperature on the top $T_{top}$ of the simulated sample provided the thermal conductivities are considered temperature-independent. The lateral and bottom sides of the domain are considered at a fixed ambient temperature $T_{ambient}$. The sample thermal conductance within such geometry $G_{sample} = Q/(T_{top} - T_{ambient})$ is computed and compared to the conductance associated with the effective thermal conductivity, known analytically to be $G_{sample} = 4k_{eff}b$ (defined as that of an equivalent-bulk thermal conductance for the same radius), determined from Suppl. Fig. 6. When the two thermal conductances are equal, this process provides the thermal contact radius $b$. We find $b \approx 4$ µm, which confirms that air heat transfer predominates.

This work can also be performed for a 395-nm silicon oxide layer, which provides the equivalent sample thermal conductance as that of the MoS$_2$/SiO$_2$/Si system. Note that we do not consider here possible partially-ballistic dissipation in contrast to Ref. [4].

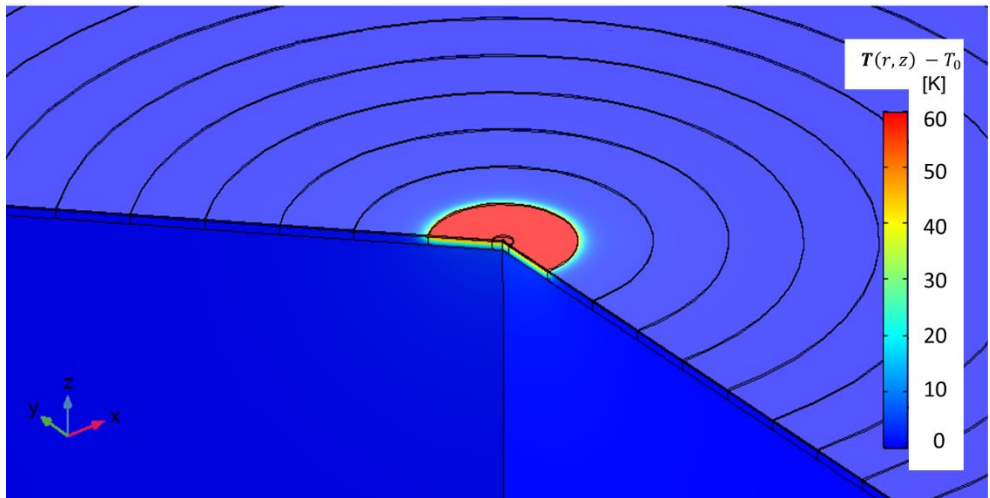

*Suppl. Fig. 7. Temperature distribution on the equivalent (395 nm SiO$_2$ film on Si substrate) sample surface.*

## 8. Determination of the thermal boundary conductance

In the final step, we perform simulations with the thermal radius previously determined by varying the thermal conductivity of MoS$_2$ $k$ and the thermal boundary conductances $G_{TBC}$ at its boundary with SiO$_2$. While the values of thermal conductivity $k$ do not impact much on the total sample thermal conductance (which is known to be $4k_{eff}b$, where $b$ is the thermal radius), one value of the boundary conductance provides the correct sample thermal conductance. The cross section temperature field is shown in Suppl. Fig. 8, and as a function of depth $z$ on the revolution axis in the core paper.

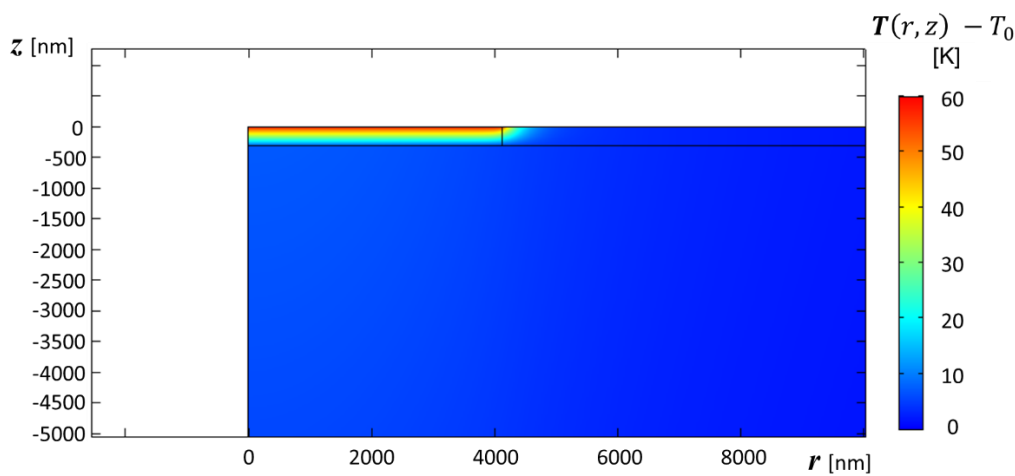

*Suppl. Fig. 8. Temperature field in a cross section of the MoS2/SiO2/Si sample.*